\documentstyle[aps,prl]{revtex}
\begin{document}
\title{ 
Beyond Poisson-Boltzmann: Fluctuations and Correlations
}

\author{
Roland R. Netz$^{\$}$ and 
Henri Orland$^*$
}
\address{$^{\$}$Max-Planck-Institut f\"ur Kolloid- und Grenzfl\"achenforschung,
Kantstr. 55, 14513 Teltow, Germany}
\address{$^*$Service de Physique Th\'eorique, 
CEA-Saclay, 91191 Gif sur Yvette, France}

\date{\today}

\maketitle

\begin{abstract}

We formulate the non-linear field theory for a fluctuating 
counter-ion distribution in the presence of a fixed,
arbitrary charge distribution. The Poisson-Boltzmann equation
is obtained as the saddle-point, and the effects of 
fluctuations and correlations are included by a loop-wise
expansion around this saddle point. We show that 
the Poisson equation is obeyed at each order
in the loop expansion and explicitly give the 
expansion of the Gibbs potential up to two loops.
We then apply our formalism to the case of an impenetrable,
charged wall, and obtain the fluctuation corrections
to the electrostatic potential  and counter-ion
density to one-loop order without further approximations. 
The relative importance
of fluctuation corrections is controlled by a single parameter,
which is proportional to the cube of the counter-ion valency
and to the surface charge density.
We also calculate effective interactions between charged particles, 
which reflect counter-ion correlation effects.

\end{abstract}

\section{Introduction}

The behavior of charged, fluctuating systems is an old problem
in chemistry and physics and is of importance for very diverse
disciplines. In this article we will be concerned with the
distribution of counter ions around charged objects, which 
is experimentally realized whenever an object with dissociable
surface groups is brought in contact with water (or some other
high-dielectric solvent). In such a situation, the counter ions
will be attracted to the oppositely charged object, but also 
repelled from other counter ions. The characteristic feature 
of charged systems is that the Coulomb interaction is long-ranged,
which gives these systems very special properties.
Since, in general, one counter ion interacts with many different
counter  ions simultaneously, the mean-field approach is very successful
and can be used to describe experiments or simulations 
quantitatively. Exact solutions of the so-called Poisson Boltzmann
equation, which determines the electrostatic potential distribution 
on a mean-field level, are available for planar\cite{Gouy,Chapman,Stern}
and cylindrical\cite{Fuoss} geometries.
A very readable introduction into the Poisson-Boltzmann (PB) approach
is given in Ref.\cite{Andelman}.
There are several factors which contribute to deviations from the
PB equation, including additional, short-ranged
interactions and  solvent effects. In this article we will 
consider deviations due to {\em fluctuations and correlations}.
These effects have been incorporated for the single charged wall
by modified PB equations \cite{Outh}, by integral equation
theories\cite{Henderson,Forstmann}, and by numerical 
methods\cite{Torrie,Megen,Svensson,Svensson2}. In general,
one finds correlation effects to become important for ions
of high valency, and packing effects dominate the ion distribution
for very bulky ions or high ion concentration.
For the case of counter ions or electrolyte solutions between
two charged walls, there have been a number of studies using
integral equations\cite{Marcelo,Kjellander,Kjellander2}
and Monte-Carlo methods\cite{Jonsson,Guldbrand}.
An exhaustive review has been given by Attard\cite{Attard}.
The main result of these studies is that correlations can lead
to an effective attraction between similarly charged walls for
high enough ion valency.
There have been  a few attempts of using field theoretic 
methods to go beyond mean-field theory for charged 
systems. Here we only mention the works  by Podgornik and Zeks\cite{Podgornik}
and by Attard and coworkers\cite{Attard2}, 
who expanded around the mean-field solution
and calculated within additional approximations  fluctuation corrections
to the interaction between two charged, planar walls.

In this article we consider a system of mobile, point-like
counter ions in the presence of a fixed charge distribution,
and formulate the field-theoretic framework 
for going beyond  the Poisson-Boltzmann description. As has been 
realized by many others before, the Poisson-Boltzmann equation
constitutes the {\em saddle point} of the exact field theory. 
We first reformulate this theory by a loop-wise
expansion around the saddle point. The Poisson equation,
which relates the electrostatic potential to the counter ion
distribution, is obeyed at each level of this loop expansion.
Clearly, the counter-ions are not distributed according to the
Boltzmann weight, and correlations and fluctuations lead to 
pronounced deviations from the Poisson-Boltzmann predictions.
The description is considerably facilitated
by a Legendre transformation, after which we obtain the Gibbs 
potential as a function of a fixed electrostatic potential distribution. 
This transformation restricts the number of diagrams  to so-called
one-particle-irreducible diagrams. The
electrostatic potential is given by the solution of the equation of state.

We then apply this general formalism to the case of counter ions
confined to a half space and next to a charged wall.
We calculate without further approximations  the one-loop correction
to the electrostatic potential distribution and the counter-ion
distribution. Due to correlations the counter ions are 
more densely packed close to the wall
than predicted by the PB solution. This is in
accord with previous Monte-Carlo simulations and
with the intuitive expectation, since mean field usually
overestimates repulsive interactions due to the neglect of 
correlations. The relative importance of fluctuation corrections
to the electrostatic potential is measured by the single parameter
$q^3 \sigma \ell_B^2$, where $\sigma$ is the surface charge density of the
planar surface, $\ell_B$ is the distance at which two elementary charges
interact with thermal energy, and $q$ denotes the counter-ion valency. 
Since the thickness of the counter ion
distribution is, according to the mean-field prediction,
given by $\mu \sim 1/(q \sigma \ell_B)$, the average 
counter-ion concentration follows as $c \sim \sigma /q \mu$ and
our effective fluctuation parameter can be rewritten as
$q^3 \sigma \ell_B^2 \sim (q^2 \ell_B c^{1/3})^{3/2}$ 
and thus measures the electrostatic interaction energy
between two counter ions at their average separation (in units of the
thermal energy); it  is thus related to the plasma parameter. 
For vanishing values of  $q^3 \sigma \ell_B^2$ mean field theory
becomes exact  and fluctuation corrections are unimportant. 
On the other hand, since the counter ion valency comes in
as a cubic power, it is clear that by going from monovalent 
to divalent ions, fluctuation effects become much more pronounced,
in accord with experiments and simulations.
Our results, which  are obtained only to first order in the
 loop expansion, 
become unreliable for large values of $q^3 \sigma \ell_B^2$. A clear
break down of our expansion is indicated by an unphysical 
negative density of counter ions,
which occurs at $q^3 \sigma \ell_B^2 \simeq 12$.

In Section II we formulate the general field-theoretic description
for an ensemble of fluctuating counter ions in the presence of 
a fixed charge distribution. 
The main steps consist of i) formulating the initial partition 
function as a field theory using a Hubbard-Stratonovich transformation,
ii) going to the grand-canonical ensemble, iii) expanding the 
action around the saddle-point, and iv)  performing another Legendre
transformation after which the Gibbs potential depends on the 
electrostatic potential. 
In Section III we show how  the electrostatic potential follows 
from the equation of state, and in Section IV we give explicit 
results for the case of a single charged wall.
In Section V  we calculate the effective interaction between
charged test particles, and Section VI is devoted to a brief discussion.

\section{Non-linear field theory for charged systems}

The partition function for $N$ mobile counter ions of valency $q$, 
interacting with an arbitrary fixed charge distribution,
$\sigma({\bf r})$, reads
\begin{equation}
\label{model1}
Z_N = \frac{1}{N!} \left[ \prod_{i=1}^N \int {\rm d} {\bf r}_i \right]
\exp\left\{ -q^2 \sum_{j > k} v({\bf r}_j 
- {\bf r}_k) + q \int {\rm d} {\bf r} \sigma({\bf r}) 
\sum_j
v({\bf r} - {\bf r}_j) \right\}.
\end{equation}
We assume the counter ions to be point-like particles,
and only include  the Coulomb interaction between ions,
$v({\bf r}) \equiv  \ell_B /r$, where $\ell_B$  is the  Bjerrum length
defined as $\ell_B \equiv e^2/(4 \pi \varepsilon k_BT)$.
We also neglect any image charge effects in the present 
formulation of our theory.
The charge distribution $\sigma({\bf r})$
in general also contains test particles,
the presence of which allows to calculate the effective interaction
between charges, as we will demonstrate in Section VI.
Introducing the particle density operator
\begin{equation}
\hat{\rho}({\bf r}) \equiv  \sum_{i=1}^N   
\delta({\bf r} -{\bf r}_i) ,
\end{equation}
 we can rewrite the partition function as
\begin{equation}
Z_N[h] = \frac{1}{N!} \left[ \prod_{i=1}^N \int {\rm d} {\bf r}_i \right]
\exp\left\{ - \frac{1}{2} \int {\rm d} {\bf r}
{\rm d} {\bf r}' [ q \hat{\rho}({\bf r}) - \sigma({\bf r})] 
v({\bf r}-{\bf r}')  [ q \hat{\rho}({\bf r}') -\sigma({\bf r}')]
+ \int {\rm d} {\bf r} \hat{\rho}({\bf r})
h({\bf r}) \right\}.
\end{equation}
There is no need to subtract the diagonal terms involving 
self interactions at this point, because they will be 
canceled by a one-point renormalization at a later stage.
Using the generating field $h$, the
average counter ion density  can be calculated as
\begin{equation}
\label{density}
\langle \hat{\rho}({\bf r}) \rangle =
\left. \frac{\delta Z_N[h] }{ Z_N[h] \delta h({\bf r})} \right|_{h=0}.
\end{equation}
Correlation functions can be calculated by taking multiple 
functional derivatives with respect to $h$.
Noting that the operator-inverse
of the Coulomb interaction is
\[
v^{-1}({\bf r}) = - \frac{\nabla^2 \delta({\bf r})}{4 \pi \ell_B},
\]
the partition function  is after a 
Hubbard-Stratonovich transformation given by
\begin{equation}
Z_N[h] = \frac{1}{N!} \int \frac{{\cal D}\phi}{Z_0} \; \exp \left\{-
\int {\rm d}{\bf r} \left[
\frac{1}{2 \tilde{\ell}_B} (\nabla \phi)^2
+\frac{\imath \sigma({\bf r}) \phi({\bf r})}{q}  \right] \right\}
\left[ \int {\rm d}{\bf r} {\rm e}^{- \imath  \phi({\bf r}) + h({\bf r})}
\right]^N, 
\end{equation}
where we introduced  the rescaled Bjerrum length  
$\tilde{\ell}_B  \equiv 4 \pi q^2 \ell_B$.
In the above expression, $Z_0$ denotes the partition function
of the Coulomb operator, $Z_0 \sim \sqrt{\det v}$.
Using the definition Eq. (\ref{density}), the average 
particle  density turns out to be
\begin{equation}
\label{density2}
\langle \hat{\rho}({\bf r}) \rangle = 
 N \left\langle \frac{{\rm e}^{- \imath \phi({\bf r})}}
{\int {\rm d}{\bf r} {\rm e}^{- \imath \phi({\bf r})}}
\right\rangle,\end{equation}
and the normalization property of the density distribution, 
$\int {\rm d} {\bf r} \hat{\rho}({\bf r}) = N$,
is self-evident.
The two-point cumulant correlation function 
(for ${\bf r} \neq {\bf r'}$) follows as
\begin{eqnarray}
\label{corr1}
\langle \hat{\rho}({\bf r}) \hat{\rho}({\bf r'})  \rangle &=&
 \left. \frac{\delta^2  Z_N[h] }{ Z_N[h] 
\delta h({\bf r}) \delta h({\bf r'})} \right|_{h=0}
\nonumber \\
&=& N(N-1) \left\langle \frac{{\rm e}^{- \imath \phi({\bf r})
- \imath \phi({\bf r'})}}
{\left[ \int {\rm d}{\bf r} {\rm e}^{- \imath \phi({\bf r})} 
\right]^2} \right\rangle, \end{eqnarray}
and  higher-order correlation function can be calculated in a 
similar manner.
The functional form of the particle distribution, Eq.(\ref{density2}), 
 shows that $\imath \phi$ is the reduced electric potential.
The partition function is brought into a more manageable form
by going to the grand-canonical ensemble,
\begin{equation}
\label{Zlambda}
Z[\varrho,h] \equiv \sum_{N=0}^{\infty} \lambda^{ N} Z_N[h] =
\int \frac{{\cal D}\phi}{Z_0}
\; \exp \left\{-\ell {\cal H}[\phi,h]
+ \imath \ell \int {\rm d}{\bf r}
\phi({\bf r}) \varrho ({\bf r}) \right\}.
\end{equation}
where $\lambda$ is the fugacity of the counter-ions, and
the Hamiltonian is defined as
\begin{equation}
\label{Ham}
{\cal H}[\phi,h] \equiv 
\int {\rm d}{\bf r} \left[ 
\frac{1}{2 \tilde{\ell}_B}
(\nabla \phi)^2 +\frac{\imath \sigma({\bf r}) \phi({\bf r})}{q}
- \lambda {\rm e}^{- \imath \phi({\bf r})+ h({\bf r})} \right]
\end{equation}
and we added 
a source term to the partition function 
in such a way that one can calculate directly
the reduced electrostatic potential
\begin{equation}
\label{defpsi}
\psi({\bf r}) \equiv
\langle \imath \phi({\bf r}) \rangle =
\frac{\delta \ln \; Z[\varrho,h] }{
\ell \delta \varrho({\bf r})}.
\end{equation}
In the grand-canonical partition function, Eq.(\ref{Zlambda}),
we arbitrarily multiplied the action by a constant $\ell$
which plays the role similar to the inverse Planck's constant in front
of the action integral in Quantum Field Theory.
This constant serves as an expansion parameter in our  systematic 
treatment of fluctuation effects and counts the number of 
diagrammatic loops.
The fugacity $\lambda$ is related to the total number of ions by
\begin{equation}
\label{lambda/N}
\lambda \frac{\partial \ln \; Z[\varrho,h] }{\ell
\partial \lambda} = \langle N  \rangle = 
\lambda \int {\rm d}{\bf r}\; \left\langle {\rm e}^{- \imath \phi({\bf r})}
\right\rangle 
\end{equation}
which establishes a useful  relation between $\lambda$ and $N$.
The particle density is obtained from the grand-canonical partition
function Eq.(\ref{Zlambda}) using the definition Eq.(\ref{density})
and reads
\[
\langle \hat{\rho}({\bf r}) \rangle =
\lambda  \left\langle {\rm e}^{- \imath \phi({\bf r})}
\right\rangle.\]
Likewise, the two-point correlation function 
(for ${\bf r} \neq {\bf r'}$) reads
\begin{equation}
\left\langle \hat{\rho}({\bf r}) \hat{\rho}({\bf r'})  
\right\rangle =
\lambda^2  \left\langle {\rm e}^{- \imath \phi({\bf r})
- \imath \phi({\bf r'})} \right\rangle, 
\end{equation}
which is simpler than the canonical form, Eq.(\ref{corr1});
higher-order correlation function can be calculated similarly.
In the following we will be interested in calculating the Gibbs potential 
$\Gamma[\psi,h]$ which depends on the  reduced electric potential
$\psi({\bf r})$;
it is related to $\ln Z[\varrho,h]$ by a Legendre transform
\begin{equation}
\label{legendre}
\Gamma[\psi,h] + \ln Z[\varrho,h]= \ell \int {\rm d}{\bf r}\; \psi({\bf r})
\varrho ({\bf r}).
\end{equation}
Using the definition (\ref{defpsi}) we obtain from the definition
of the Legendre transform (\ref{legendre}) the inverse relation
\begin{equation} \label{varrho}
\varrho({\bf r}) =
\frac{\delta \Gamma[\psi,h]}{\ell \delta \psi({\bf r})}.
\end{equation}
We will perform the Legendre transformation using a loop expansion,
using methods developed by Schwinger\cite{Schwinger}
and closely follow the notation of 
Br\'ezin, Le Guillou, and Zinn-Justin\cite{BLZ}.
To that end, we expand the Hamiltonian around a fixed 
value of the potential, the so-called classical field, denoted
as $\psi_{\rm cl}({\bf r})$.
At this point, we do not specify what this classical field is.
We write 
\begin{equation}
\label{sep}
\phi({\bf r}) = - \imath \psi_{\rm cl}({\bf r}) +\chi({\bf r})
\end{equation}
and thus obtain the formal expansion 
\[
{\cal H}[\phi]
- \imath \int {\rm d}{\bf r}
\phi({\bf r}) \varrho ({\bf r}) = 
{\cal H}[- \imath \psi_{\rm cl}] -\int {\rm d}{\bf r}
\varrho ({\bf r})\left[\psi_{\rm cl}({\bf r})
+\imath\chi({\bf r}) \right]
+\sum_{j=1}\frac{1}{j!} \int {\cal H}^{(j)}(\{{\bf r}_j\}; \psi_{\rm cl})
\chi({\bf r}_1)\cdots \chi({\bf r}_j) {\rm d}{\bf r}_1 \cdots
{\rm d}{\bf r}_j.\]
The vertex functions ${\cal H}^{(j)}$ are defined by
\begin{equation}
\label{vertdef}
{\cal H}^{(j)}(\{{\bf r}_j\}; \psi_{\rm cl}) \equiv \left.
\frac{\delta^j {\cal H}[ \phi]}
{\delta \phi({\bf r}_1)\cdots \delta \phi({\bf r}_j)}
\right|_{\phi=-\imath \psi_{\rm cl}}
\end{equation}
and of course depend on the function $\psi_{\rm cl}({\bf r})$ 
around which one expands.
The function $\ln Z[\varrho]$ can now be evaluated 
by standard saddle-point
methods, treating $\ell^{-1}$ as the expansion parameter.
The saddle point $\psi_{\rm SP}({\bf r})$
is defined by
\begin{equation}
\label{saddle}
\left.
\frac{ \delta {\cal H}[\phi]}{\delta \phi({\bf r})} -
\imath \varrho({\bf r}) \right|_{\phi=-\imath \psi_{\rm SP}[\varrho]}=0
\end{equation}
and depends on the source $\varrho$.
Using the definition of the two-point propagator 
$G({\bf r},{\bf r}';\psi_{\rm SP})$,
\[
\int {\rm d}{\bf r} \;  G({\bf r},{\bf r}';\psi_{\rm SP})
{\cal H}^{(2)}({\bf r},{\bf r}'';\psi_{\rm SP}) 
= \delta({\bf r}'-{\bf r}''),
\]
we can diagrammatically expand the expression for 
$\ln Z[\varrho,h]$ and obtain up to two loops
\begin{eqnarray} \label{logZ}
\ln Z[\varrho,h] &=&
\ell \int {\rm d}{\bf r} \left\{ \frac{1}{2 \tilde{\ell}_B}
(\nabla \psi_{\rm SP})^2 -\frac{\sigma({\bf r}) \psi_{\rm SP}({\bf r})}{q} 
+ \lambda {\rm e}^{-  \psi_{\rm SP}({\bf r})+h({\bf r})}  
+ \varrho ({\bf r})\psi_{\rm SP}({\bf r}) \right\}
- \frac{1}{2} \ln \det {\cal H}^{(2)}[\psi_{\rm SP}] \nonumber \\
&+& \frac{1}{\ell} \left\{ -\frac{1}{8} \int {\rm d}{\bf r}_1 
\cdots {\rm d}{\bf r}_4 \; {\cal H}^{(4)}(\{{\bf r}_4\};\psi_{\rm SP})
G({\bf r}_1,{\bf r}_2;\psi_{\rm SP}) 
G({\bf r}_3,{\bf r}_4;\psi_{\rm SP}) \right. \nonumber \\
&&\;\;\;\;\;\;\left. +\int {\rm d}{\bf r}_1\cdots {\rm d}{\bf r}_3
{\rm d}{\bf r}'_1\cdots {\rm d}{\bf r}'_3
\; {\cal H}^{(3)}(\{{\bf r}_3\};\psi_{\rm SP})
{\cal H}^{(3)}(\{{\bf r}'_3\};\psi_{\rm SP}) \right. \nonumber \\
&& \left. \left[ 
\frac{1}{8} G({\bf r}_1,{\bf r}_2;\psi_{\rm SP})
G({\bf r}'_1,{\bf r}'_2;\psi_{\rm SP})
G({\bf r}_3,{\bf r}'_3;\psi_{\rm SP})
+\frac{1}{12} G({\bf r}_1,{\bf r}'_1;\psi_{\rm SP})
G({\bf r}_2,{\bf r}'_2;\psi_{\rm SP})
G({\bf r}_3,{\bf r}'_3;\psi_{\rm SP}) \right] \right\}.
\end{eqnarray}
The one-loop diagram and the two-loop diagrams are schematically
represented in Fig.1. 
In order to perform the Legendre transformation, we need the
loop expansions for $\psi({\bf r})$ and $\varrho({\bf r})$,
which are obtained from  the definition (\ref{defpsi}) and
the saddle-point equation (\ref{saddle}), respectively.
These expressions depend on the saddle-point potential $\psi_{\rm SP}$.
We then have to expand all functions and vertices around the saddle point
and reexpress the potential dependence as a function of
the expectation value $\psi$. By inserting
these expression into the Legendre transform,
we obtain the loop expansion of the Gibbs potential,
\begin{eqnarray} \label{Gamma}
\Gamma[\psi,h] &=&
-\ell \int {\rm d}{\bf r} \left\{ 
\frac{1}{2 \tilde{\ell}_B}
(\nabla \psi)^2 - \frac{\sigma({\bf r}) \psi({\bf r})}{q} + \lambda  
{\rm e}^{-  \psi({\bf r})+h({\bf r})}  \right\}
+ \frac{1}{2} \ln \det {\cal H}^{(2)}[\psi] \nonumber \\
&-& \frac{1}{\ell} \left\{ -\frac{1}{8} \int {\rm d}{\bf r}_1 
\cdots {\rm d}{\bf r}_4 \; {\cal H}^{(4)}(\{{\bf r}_4\};\psi)
G({\bf r}_1,{\bf r}_2;\psi) 
G({\bf r}_3,{\bf r}_4;\psi) \right. \nonumber \\
&&\;\;\;\;\;  \left. +\frac{1}{12} \int {\rm d}{\bf r}_1\cdots {\rm d}{\bf r}_3
{\rm d}{\bf r}'_1\cdots {\rm d}{\bf r}'_3
\; {\cal H}^{(3)}(\{{\bf r}_3\};\psi)
{\cal H}^{(3)}(\{{\bf r}'_3\};\psi) \right. \nonumber \\
&& \left. \; \;\;\;\;\;\;
G({\bf r}_1,{\bf r}'_1;\psi)
G({\bf r}_2,{\bf r}'_2;\psi)
G({\bf r}_3,{\bf r}'_3;\psi) \right\}
\end{eqnarray}
The effect of the Legendre transformation is to cancel all 
one-particle-reducible diagrams, as expected. For the present
case, the two-loop diagram to the right  in Fig.1 is removed from
the expansion.

\section{The equation of state}
%
%
The equation of state is defined as
\begin{equation}
\frac{\delta \Gamma[\psi]}{\delta \psi({\bf r})}=0
\end{equation}
and completely determines the electrostatic potential.
In the following, we will restrict ourselves to the one-loop order.
Using the explicit form of the two-point vertex function,
which follows from the definitions Eq.(\ref{Ham}) and Eq.(\ref{vertdef}),
\begin{equation}
{\cal H}^{(2)}[{\bf r},{\bf r'};\psi]=
\left( -\frac{\nabla^2}{\tilde{\ell}_B}
+\lambda {\rm e}^{-\psi({\bf r})} \right) \delta({\bf r}-{\bf r'}),
\end{equation}
the equation of state can be explicitly written as 
\begin{equation} \label{stateeq}
-\lambda {\rm e}^{-\psi({\bf r})}-\frac{\sigma({\bf r})}{q}
-\frac{\nabla^2 \psi({\bf r})}{\tilde{\ell}_B}
+\frac{\lambda}{2\ell} {\rm e}^{-\psi({\bf r})} 
G({\bf r},{\bf r}) =0.
\end{equation}
Note that from now one, we suppress the dependence of the
two-point correlation function $G({\bf r},{\bf r'})$ on the electrostatic
potential distribution $\psi$.
The correlation function is determined by the equation
\begin{equation}
\left( -\frac{\nabla^2}{\tilde{\ell}_B}
+\lambda {\rm e}^{-\psi({\bf r})} \right) 
G({\bf r},{\bf r'})=
\delta({\bf r}-{\bf r'}).
\end{equation}
Finally, the number of mobile ions can be calculated according
to Eq.(\ref{lambda/N}) and is given by
\begin{equation} \label{expectN}
\langle N \rangle = 
\lambda \int {\rm d}{\bf r}\; {\rm e}^{-\psi({\bf r})}-
\frac{\lambda}{2 \ell} \int {\rm d}{\bf r} \; 
{\rm e}^{-\psi({\bf r})} G({\bf r},{\bf r}).
\end{equation}
The equation of state can be solved by a systematic expansion of
all quantities which are to be determined
in inverse powers of the loop parameter $\ell$.
At the one-loop order, only the leading term of $G$
contributes and therefore an expansion of $G$ itself is unnecessary.
The electric potential and the fugacity have to be expanded 
to the next-leading order,
\begin{equation}
\label{psiexp}
\psi({\bf r})=\psi_0({\bf r}) + \ell^{-1} \psi_1({\bf r}),
\end{equation}
\begin{equation}
\lambda=\lambda_0 + \ell^{-1} \lambda_1.
\end{equation}
The equation of state splits into two separate equations,
the zero-loop equation
\begin{equation}
\label{psi0def}
-\lambda_0 {\rm e}^{-\psi_0({\bf r})}-\frac{\sigma({\bf r})}{q}
-\frac{\nabla^2 \psi_0({\bf r})}{\tilde{\ell}_B} =0,
\end{equation}
which is the ordinary Poisson Boltzmann equation, and the
next-leading correction,
\begin{equation}
\label{psi1def}
-\lambda_1 {\rm e}^{-\psi_0({\bf r})}
+\lambda_0 {\rm e}^{-\psi_0({\bf r})} \psi_1({\bf r})
-\frac{\nabla^2 \psi_1({\bf r})}{\tilde{\ell}_B}
+\frac{\lambda_0}{2} {\rm e}^{-\psi_0({\bf r})} 
G({\bf r},{\bf r}) =0.
\end{equation}
The correlation function is determined by
\begin{equation}
\label{Gdef}
\left( -\frac{\nabla^2}{\tilde{\ell}_B}
+\lambda_0 {\rm e}^{-\psi_0({\bf r})} \right) 
G({\bf r},{\bf r'})=
\delta({\bf r}-{\bf r'}).
\end{equation}
The fugacity is at the zero-loop level determined by
\begin{equation}
\label{lambda0}
\lambda_0 = 
\frac{\langle N \rangle}{
\int {\rm d}{\bf r}\; {\rm e}^{-\psi_0({\bf r})}}
\end{equation}
and at the next-leading level by
\begin{equation}
\label{lambda1a}
\lambda_1 = 
\lambda_0 \frac{ \int {\rm d}{\bf r}\; {\rm e}^{-\psi_0({\bf r})}
\psi_1({\bf r})}{\int {\rm d}{\bf r}\; {\rm e}^{-\psi_0({\bf r})}}
+\frac{\lambda_0}{2} \frac{ \int {\rm d}{\bf r}\; {\rm e}^{-\psi_0({\bf r})}
G({\bf r},{\bf r})}{
\int {\rm d}{\bf r}\; {\rm e}^{-\psi_0({\bf r})}}.
\end{equation}
Combining the equations for $\psi_1$ and $G$,
Eqs.(\ref{psi1def}) and (\ref{Gdef}), we 
obtain for the correction to the electrostatic potential
\begin{equation}
\label{psi1a}
\psi_1({\bf r}) = -\frac{\lambda_0}{2} \int {\rm d} {\bf r}' \;
{\rm e}^{-\psi_0({\bf r}')} G({\bf r}',{\bf r})
\left[ G({\bf r}',{\bf r}') - \frac{2 \lambda_1}{\lambda_0}
\right],
\end{equation}
which contains an implicit dependence on $\psi_1$ on the right-hand
side through the dependence of $\lambda_1$ on $\psi_1$, see
Eq.(\ref{lambda1a}). Algebraically solving
Eqs.(\ref{lambda1a}) and (\ref{psi1a}) we obtain
\begin{equation}
 \int {\rm d}{\bf r}\; {\rm e}^{-\psi_0({\bf r})}
\psi_1({\bf r}) = \frac{-(\lambda_0/2) \int {\rm d}{\bf r}
{\rm d}{\bf r'}\; {\rm e}^{-\psi_0({\bf r})-\psi_0({\bf r'})}
G({\bf r}',{\bf r}) \left[ G({\bf r},{\bf r})-Z/Y \right]}
{1-(\lambda_0/Y) \int {\rm d}{\bf r} {\rm d}{\bf r'}\;
 {\rm e}^{-\psi_0({\bf r})-\psi_0({\bf r'})}
G({\bf r}',{\bf r})},
\end{equation}
where we used the short-hand notations
\begin{equation}
Y =  \int {\rm d}{\bf r}\; {\rm e}^{-\psi_0({\bf r})}
\end{equation}
and 
\begin{equation}
Z =  \int {\rm d}{\bf r}\; {\rm e}^{-\psi_0({\bf r})}
G({\bf r},{\bf r}).
\end{equation}
 The one-loop correction to the electrostatic potential is
 therefore determined by Eq.(\ref{psi1a}) in conjunction
 with the equations determining $G$ and $\lambda_1$,
 Eqs.(\ref{Gdef}) and (\ref{lambda1a}).

\section{Single charged impenetrable  wall}

In the following we consider the case of a charged wall
which is impenetrable for the counterions. 
The fixed charge distribution is given by $\sigma({\bf r}) = 
\sigma \; \delta (z)$ where $\sigma$ is the surface charge density
at the charged wall.
The solution for the one-dimensional Poisson-Boltzmann equation,
Eq.(\ref{psi0def}), is 
\begin{equation}
\label{psi0}
\psi_0(z)= 2 \ln(1+z/\mu),
\end{equation}
where $\mu =  \sqrt{ 2/ \tilde{\ell}_B \lambda_0} = 1/ \sqrt{2 \pi q^2 \ell_B \lambda_0}$ 
is the Gouy-Chapman
length. Enforcement of  the normalization condition,  Eq.(\ref{lambda0}),
leads to a relation between
the surface charge density $\sigma$ and the Gouy-Chapman length,
$\sigma = q \lambda_0 \mu$,  which can be further
transformed to $\lambda_0 = 2 \pi \ell_B 
\sigma^2$ and $1/ \mu = 2 \pi \ell_B q \sigma$, the standard results.

\subsection{Calculation of the correlation function}

Due to lateral invariance,
the Green's function  $G({\bf r},{\bf r}')$
can be Fourier transformed in the directions
perpendicular to the surface; the equation for the Green's function,
Eq.(\ref{Gdef}), becomes
\begin{equation}
\left( -\frac{{\rm d}^2}{{\rm d} z^2} +p^2
+\tilde{\ell}_B \lambda_0 {\rm e}^{-\psi_0(z)} \right) 
G(z,z',p)= \tilde{\ell}_B  
\delta(z-z').
\end{equation}
where $\vec p$ are the Fourier variables in the lateral directions,
and $p= |\vec p|$.
The potential $\psi_0(z)$ is given by Eq.(\ref{psi0}) for
$z>0$. Since the ions cannot penetrate into the wall, the
counter-ion density is zero for $z<0$, or, 
equivalently, we put $\psi_0(z) = \infty$ for $z<0$. 
The solution can be obtained in a straightforward
manner and reads
\begin{equation}
\label{Gsol}
G(z, z',p) = \frac{\tilde{\ell}_B}{2 p^5}
\left( \frac{p}{z'+\mu}+p^2 \right)
\left[ \left( -\frac{p}{z+\mu}+p^2 \right) e^{p (z-z')}
+ \left( \frac{ \frac{\displaystyle p}{\displaystyle z+\mu}+p^2 }
{1+2 p^2 \mu^2 + 2 p \mu} \right) e^{-p (z+z')} \right]
\end{equation}
for $z' > z >0 $.  
The solution for $z>z'>0 $  follows from Eq.(\ref{Gsol}) 
by an interchange of the arguments $z$ and $z'$.
In the limit $p = 0 $ we obtain
\begin{equation}
G(z, z',p=0) = 
\frac{\tilde{\ell}_B \mu  (2  + (1+z/\mu)^3)}
{3 (1+z/\mu)(1+z'/\mu)}
\end{equation}
for $z' > z >0 $ (and with the coordinates interchanged
for $z>z' >0$).
The self interaction is defined as the equal-point
Green's function and follows from the partially Fourier-transformed
Green's function as
\begin{equation}
\label{Gselfdef}
G_{\rm self}(z) = \int \frac{d^2p}{(2\pi)^2}
G(z,z,p) .
\end{equation}
The integral to be solved is, subtracting the 
(infinite) Coulomb self energy $G_{\rm self}(z = \infty)$,
\begin{equation}
G_{\rm self}(z) = \frac{\tilde{\ell}_B}{2} \int_0^\infty
\frac{d p}{2 \pi}  \; \left[ -\frac{1}{(z+\mu)^2 p^2}
+\frac{1+2p^{-1} (z+\mu)^{-1} +p^{-2} (z+\mu)^{-2}}
{1+2 p^2 \mu^2 + 2 p \mu}e^{-2p z} \right].
\end{equation}
The integral can be performed by some intermediate transformations,
and the result can be written in a scaling form as 
\begin{equation}
\label{Gselfsol}
G_{\rm self}(z) = \frac{\tilde{\ell_B}}{4 \pi \mu} g(z/\mu)
\end{equation}
where the scaling function $g(x)$ is explicitly given by
\begin{equation}
g(x)= \frac{1}{2 (1+x)^2}  \left[
i e^{(1-i)x} E_1[(1-i)x](1+ix)^2 -
i e^{(1+i)x} E_1[(1+i)x](1-ix)^2 - 4x \right]
\end{equation}
and plotted in Fig.2.
The function $E_1[x]$ is the exponential-integral 
function\cite{Abramowitz}. 
The prefactor of the scaling function in Eq.(\ref{Gselfsol}),
\begin{equation}
\frac{\tilde{\ell_B}}{4 \pi \mu} =
\frac{q^2 \ell_B} {\mu} =  2 \pi q^3 \ell_B^2 \sigma,
\end{equation}
is a measure of the importance of correlation effects and
thus the departure from mean-field behavior. We note that
it is this single parameter combination which measures
the relative importance of fluctuation corrections. Clearly,
the cubic dependence on the counter-ion valency shows that
fluctuation effects will become much more dominant for multivalent
ions. As can be seen from Fig.2,
the self energy is negative, which reflects the fact that the
screening due to counter ions lowers the local free energy of
charged particles. The scaling function $g(z/\mu)$  shows
a minimum at a finite distance from the wall, $z/\mu \approx 0.2$.
This maximum in screening is closely connected with a maximum 
in the counter-ion density at about this distance from the wall,
as we will show in Section IV.C.  The asymptotic behavior for
small separations from the wall is 
\begin{equation} \label{gasym1}
g(x) \simeq - \frac{\pi}{4} \left( 1 - x \ln x \right) 
\end{equation}
and 
\begin{equation} \label{gasym2}
g(x) \simeq - \frac{3}{2 x}
\end{equation}
for large separations. 
The functions Eqs.(\ref{gasym1}) and
(\ref{gasym2}) are shown in Figs.2a and 2b as broken lines.

\subsection{Calculation of the one-loop correction to the electrostatic potential}

The solution of the equation for $\psi_1$ is considerably simplified
by the fact that, for the specific mean-field potential given in
Eq.(\ref{psi0}), we find
\begin{equation}
\int {\rm d} z\; G(z,z',p=0) e^{-\psi_0(z)} = \tilde{\ell}_B \mu^2/2
\end{equation}
independent of $z'$. It immediately follows from
Eq.(\ref{psi1a}) that the integral
$\int {\rm d}z e^{-\psi_0(z)} \psi_1(z) $ is zero
and, as a consequence, the equation for
$\psi_1$, Eq.(\ref{psi1a}), is independent of 
any constant appearing in the correlation function
$G$. This holds in particular for the 
Coulombic self energy and therefore justifies that we neglected this
(infinite) constant in the calculation of the self interaction.
This is an example of a 
renormalization of a one-point correlation function,
which otherwise is plagued by the (infinite) Coulombic
self-energy.
The normalization of the bare potential $\psi_0$, Eq.(\ref{psi0})  is
\begin{equation}
\int {\rm d} z\; {\rm e}^{-\psi_0(z)} = \mu.
\end{equation}
The renormalization constant appearing in Eq.(\ref{psi1a}) thus
is
\begin{equation}
\frac{2 \lambda_1}{\lambda_0} = 
\frac{ \int {\rm d}{z}\; {\rm e}^{-\psi_0(z)}
G_{\rm self}(z)}
{ \int {\rm d}{z}\; {\rm e}^{-\psi_0(z)}} = 
-\frac{\tilde{\ell_B} c_1}{4 \pi \mu}
\end{equation}
where the numerical constant is $c_1 = \int_0^\infty {\rm d}x g(x)(1+x)^{-2} 
= 0.6208593$.
Our final result for the potential is, according to Eq.(\ref{psi1a}),
\begin{equation}
\label{psi1}
\psi_1(z) = \frac{\tilde{\ell_B} }{4 \pi \mu} 
\left[  f(z/\mu)-\frac{c_1}{2} \right],
\end{equation}
where the scaling function $f(x)$ is defined by
\begin{equation}
f(x) = - \int_0^x {\rm d} x' \frac{2+(1+x')^3}
{3(1+x)(1+x')^3} g(x') -
\int_x^\infty {\rm d} x' \frac{2+(1+x)^3}
{3(1+x)(1+x')^3} g(x')
\end{equation}
and graphically presented in Fig.3. The asymptotic behavior for small 
separations is
\begin{equation} \label{fasym1}
f(x) \simeq  \frac{\pi}{8} \left( 1 - \frac{x^3 \ln x}{3}  \right) 
\end{equation}
and for large separations we obtain
\begin{equation} \label{fasym2}
f(x) \simeq  \frac{\ln x}{2 x}.
\end{equation}
The asymptotic formulas are shown in Fig.3 as broken lines. 

\subsection{Calculation of the counter-ion density}

The counter ion density 
$\rho({\bf r})$ can be calculated from the Gibbs potential Eq.(\ref{Gamma})
by taking a derivative with respect to the generating field $h$.
The result is, including the one-loop correction, given by
\begin{equation}
\rho({\bf r}) = 
\lambda  {\rm e}^{-\psi({\bf r})}-
\frac{\lambda}{2 \ell}  
{\rm e}^{-\psi({\bf r})} G({\bf r},{\bf r})
\end{equation}
and leads via integration to the total ion
number $N$, Eq.(\ref{expectN}), as it should.
Subtracting the equation of state, Eq.(\ref{stateeq}), we obtain 
the Poisson equation,
\begin{equation} \label{Poisson}
\rho({\bf r}) = -\frac{\sigma({\bf r})}{q}- \frac{\nabla^2 \psi({\bf r})}
{\tilde{\ell}_B}.
\end{equation}
In fact, the equation Eq.(\ref{Poisson}) is exact to all orders
in the loop expansion, since the equation of state contains
exactly the same terms as does the particle density and thus leads
to an exact cancelation of all higher-loop terms. This is so
because in the Hamiltonian Eq.(\ref{Ham}), the generating 
field $h$ enters in the same way as the fluctuating potential
$\phi$, except for the linear and quadratic terms.
Expanding the counter ion density in powers of the
loop parameter,
\begin{equation}
\rho({\bf r}) = \rho_0({\bf r}) +
\ell^{-1} \rho_1({\bf r}),
\end{equation}
and using the
loop-wise expansions of the fugacity $\lambda$
and the electrostatic potential 
$\psi({\bf r})$ introduced in Section III,
the zero-loop (or mean-field) result for the counter ion density is 
\begin{equation}
\rho_0({\bf r}) = \lambda_0 {\rm e}^{-\psi_0 ({\bf r})}
\end{equation}
and the one-loop correction reads 
\begin{equation}
\rho_1({\bf r}) = \lambda_0 {\rm e}^{-\psi_0 ({\bf r})}
\left( \frac{\lambda_1}{\lambda_0} - \psi_1({\bf r})  - 
\frac{1}{2} G({\bf r},{\bf r}) \right).
\end{equation}
Using the explicit solution for the single charged wall,
Eq.(\ref{psi0}),  and the result for the mean-field
fugacity, $\lambda_0 = 2/ \tilde{\ell}_B \mu^2$,
the zero-loop density can be written as
\begin{equation}
\rho_0(z ) = \left( \frac{2}{\tilde{\ell}_B \mu^2} \right)
\frac{1}{(1+z/\mu)^2}
\end{equation}
and the one-loop correction reads 
\begin{equation}
\label{hdef}
\rho_1(z ) = \left( \frac{2}{\tilde{\ell}_B \mu^2} \right)
\frac{\tilde{\ell}_B}{4 \pi \mu} h(z/\mu)
\end{equation}
where the scaling function $h(x)$ can be written  in terms of 
the previously defined scaling functions and reads
\begin{equation}
\label{hsol}
h(x) = - \frac{ f(x) + g(x)/2}{(1+x)^2} .
\end{equation}
In Fig.4 we plot the scaling function $h(x)$.  
The density correction vanishes at the charged wall,
and for small separations the asymptotic behavior follows from
Eqs. (\ref{gasym1}) and (\ref{fasym1}) as
\begin{equation} \label{hasym1}
h(x) \simeq - \frac{\pi}{8} x \ln x.
\end{equation}
The density corrections is positive for $x<1$, i.e. for distances
from the wall smaller than the Gouy-Chapman length $\mu$ fluctuations
and correlations lead to an enhanced density, quite in accord with
expectations: Each counter ion is surrounded by a correlation hole,
which is neglected on a mean-field level. Mean field theory therefore
overestimates the repulsion between counter ions, and therefore 
underestimates the density close to the charged wall. This is in
qualitative agreement with Monte-Carlo simulations\cite{Jonsson}.
We note that since the slope of the density correction is infinite 
close to the wall, see Eq.(\ref{hasym1}), our results predict
always an initial increase of the counter-ion density as one moves
away from the substrate.
The total integral over the density correction vanishes; the increase
in density close to the wall is therefore compensated by a decrease
in density further away from the wall, as shown in Fig.4b.
For large separations from the wall,
the asymptotic behavior follows from
Eqs. (\ref{gasym1}) and (\ref{fasym1}) as
\begin{equation}
h(x) \simeq - \frac{\ln x}{x^3}.
\end{equation}
In Fig. 5 we plot the rescaled density 
$\bar{\rho}(z) = (\tilde{\ell}_B \mu^2/2) \rho(z)$ 
as a function of the rescaled distance from the wall
$z/\mu$, for four different values of the parameter
$\tilde{\ell}_B/4 \pi \mu$. The broken line, for 
$\tilde{\ell}_B/4 \pi \mu = 0$, denotes the PB result,
the other three solid lines  are for $\tilde{\ell}_B/4 \pi \mu =1$,
$5$, and $10$, with the difference to the broken  line
increasing as the parameter $\tilde{\ell}_B/4 \pi \mu$ 
increases. We see that already for rather small values of
$\tilde{\ell}_B/4 \pi \mu$, the
one-loop result for the counter ion distribution is quite different
from the PB result.
The density depression far away from the wall is proportional to the
parameter $\tilde{\ell}_B/4 \pi \mu$. It is clear that for very 
large values of this parameter, the sum of the zero-loop and the 
one-loop densities will become negative. This in fact happens
at $\tilde{\ell}_B/4 \pi \mu \approx 12$, which clearly sets an upper
limit to the applicability of the present one-loop calculation.
It is likely that the one-loop calculation becomes quantitatively
inaccurate even for smaller values of this parameter, but the precise
value of $\tilde{\ell}_B/4 \pi \mu$ when this happens can only
be judged by a numerical calculation (such as Monte Carlo) or
by a higher-loop calculation.

\section{Interaction between charged particles}
In this section we calculate the effective
interaction between charged test particles. As mentioned
before, test particles can be incorporated into
the model by adding them to the charge distribution in 
Eq.(\ref{model1}) according to
\begin{equation}
\sigma({\bf r}) \rightarrow \sigma({\bf r}) + 
\sum_{j=1}^{M} Q_j \delta({\bf r}-{\bf R}_j),
\end{equation}
where we have considered a collection of $j=1,\ldots,M$ 
test particles with charges $Q_j$ and positions ${\bf R}_j$.
Inserting this shifted charge distribution into the
grand-canonical partition function Eq. (\ref{Zlambda}),
the reduced test-particle free energy  turns out to be
\begin{equation}
f(\{ {\bf R}_M \}) =- \ln \left\langle {\rm e}^{- \imath
\sum_j Q_j \phi({\bf R}_j)/q} \right\rangle.
\end{equation}
Using the separation of the fluctuating field $\phi$
into the expectation value $\psi$ and the fluctuations
around the expectation value, Eq.(\ref{sep}), the free
energy can be reexpressed as 
\begin{equation}
f(\{ {\bf R}_M \}) = \sum_{j=1}^M Q_j \psi({\bf R}_j)/q
+\frac{1}{2} \sum_{j,k=1}^M Q_j Q_k G({\bf R}_j,{\bf R}_k)/q^2.
\end{equation}
Using the loop-expansion of the electrostatic potential, $\psi$,
Eq.(\ref{psiexp}), and the definition of the self interaction,
$G_{\rm self}$, Eq.(\ref{Gselfdef}), the free energy
becomes on the one-loop level
\begin{equation}
f(\{ {\bf R}_M \}) = \sum_{j=1}^M \left\{ \frac{Q_j}{q} \left[
\psi_0(Z_j)+\psi_1(Z_j)\right] 
+\frac{Q_j^2}{2q^2} G_{\rm self}(Z_j) \right\}
+\sum_{j<k}^M \frac{Q_j Q_k}{q^2} G({\bf R}_j,{\bf R}_k)
\end{equation}
and thus separates into a single-particle part and a
two-particle part. The single-particle contribution contains
the ordinary Gouy-Chapman potential, $\psi_0(z)$,
given by Eq.(\ref{psi0}),
a contribution due to fluctuation-induced changes of the
electrostatic potential, $\psi_1(z)$, 
given by Eq.(\ref{psi1}), and a contribution 
due to correlations between counterions and a test particle,
$Q_j^2 G_{\rm self}(z)$, given by Eq.(\ref{Gselfsol}). 
Since the correlation contribution
is attractive and proportional to the square of the test-particle
charge, it will be dominant for large test-particle charge
and lead to attraction even in the case of a similarly charged
wall. This somewhat
surprising behavior has in fact been seen in Monte-Carlo
simulations\cite{Svensson}.
 We note that on the two-loop level, in addition a 
three-point interaction appears.

Finally we will present results for the interaction between two particles
in the neighborhood of the charged wall and thus under the influence
of the loosely bound counter-ion cloud. To make the results somewhat
more transparent, we will confine ourselves to  particles which have the
same vertical distance from the wall. The effective interaction then follows 
by Fourier transformation and reads
\begin{equation}
G(z,r) = \int \frac{{\rm d}p p }{2 \pi} G(z,z,p) {\cal J}_0(p r) 
\end{equation}
where the Green's function $G(z,z,p)$ is determined by Eq.(\ref{Gsol})
and ${\cal J}_0(x)$ denotes the Bessel function 
of the first kind\cite{Abramowitz}.
For particles which are very close to each other, $r< \mu$, and/or
very far apart from the charged surface, $z>r$, the interaction
is basically unscreened and given by 
\begin{equation}
\label{G0}
G(z,r) \simeq \ell_B q^2 /r.
\end{equation}
For particles that are far apart from each other, $r>\mu$,
but close to the surface, $z<\mu$, the effective interaction is given
by 
\begin{equation}
\label{G1}
G(z,r) \simeq \frac{2 \ell_B q^2 \mu^2 }{r^3} \left( 1-4 \frac{z}{\mu} 
+ 6 \frac{z^2}{\mu^2} + \cdots \right).
\end{equation}
The main feature is that the effective interaction is screened and thus
reduced compared to the bare Coulomb interaction, but the screening 
is much weaker than for example in a salt solution and the interaction
decays as an inverse cube of the distance. This behavior has 
previously  been predicted by 
liquid-state calculations and is interpreted as the effective dipole-dipole
interaction between the test-charges and their associated counter-ion 
clouds\cite{Jancovici,Baus,Carnie}. 
The effective interaction between two induced dipoles is unscreened
because the lower half-space  is assumed to be
impenetrable to ions and thus allows 
the unattenuated passage of electric field lines.
From Eq.(\ref{G1}) it follows  that the screening
is maximal at a finite distance away from the wall. This is connected
with the maximum in the counter-ion distribution, see Fig.4. 
For particles that are far apart from each other, $r>\mu$,
and also relatively far apart from the surface, $\mu< z< r$, 
the effective interaction is given by 
\begin{equation}
\label{G2}
G(z,r) \simeq \frac{2 \ell_B q^2 z^2 }{r^3} 
\end{equation}
and thus increases gradually as one move away from the surface.
At a separation from the surface which equals the interparticle
distance, $z \sim r$, the effective interaction has the same 
magnitude as the bare Coulomb interaction, and we find 
the effective interaction to cross over smoothly to the bare
interaction as given by Eq.(\ref{G0}).

\section{Discussion}

In this article we have formulated the non-linear field theory 
for mobile counter ions  under the influence of fixed
charge distributions. We showed explicitly that the Poisson-Boltzmann
equation corresponds to the saddle-point equation, and how
correlations and fluctuations can be accounted for by a loop-wise
expansion of the action around this saddle point. We find
the Poisson equation to be satisfied at each order in this expansion.
Clearly, the Boltzmann equation is not satisfied when going
beyond the saddle point, and this is the reason for deviations
from the Poisson-Boltzmann equation.
Particularly useful is the introduction of the Gibbs potential
$\Gamma[\psi]$, which 
only contains one-particle irreducible diagrams (and thus reduces
the number of diagrams to be considered) and which allows to directly
calculate the electrostatic potential distribution via the equation
of state. We applied our formalism to the case of a charged wall
which is impenetrable to counter ions and obtained, to one-loop
order and without further approximations, the electrostatic potential, 
the counter-ion distribution, 
and the effective interactions between test particles. 
We find that the counter-ion density is increased close to the
wall as compared to the mean-field (Poisson-Boltzmann, PB) 
solution, however, right at the wall the density is unchanged
as compared to PB. The increase close to the wall is due to correlations
between counter ions, which are not captured in the PB approach and
which reduce the repulsion between counter ions.
 The surface potential is increased
at the surface. The relative strength of the one-loop correction
is proportional to the single parameter $\tilde{\ell}_B /4 \pi \mu = 
2 \pi q^3 \ell_B^2 \sigma $ and thus is proportional to the cube
of the counter ion valency. Our results become unphysical 
at  $\tilde{\ell}_B /4 \pi \mu \approx 12$. The value  of 
$\tilde{\ell}_B /4 \pi \mu $ up to which our results are accurate could only
be inferred from quantitative comparisons with numerical simulations, which 
we plan to do in the future. Our formalism can in principle 
applied to all situations where the PB equation can be solved
in closed-form. It would be particularly useful to obtain the
one-loop correction for a charged wall in a salt solution, 
because this would represent the bridge between
Debye-H\"uckel and Poisson-Boltzmann theory.

\begin{figure}
\caption{ a) One-loop and b) two-loop diagrams that enter the
calculation of the logarithm of the partition function, 
Eq.(\ref{logZ}). The two-loop diagram to the right is one-particle
reducible and thus cancels out when going to the Gibbs potential,
Eq.(\ref{Gamma}). }
\caption{Plot of the rescaled self-energy $g$ 
as a function of the rescaled distance from the wall, $z/\mu$, as
determined by Eq.(\ref{Gselfsol}). The asymptotic behavior for 
small and large wall distance, Eqs.(\ref{gasym1}) and (\ref{gasym2}),
are drawn as broken lines in a) and b), respectively.}
\caption{Plot of the rescaled correction to the electrostatic
potential $f$,
as defined by Eq.(\ref{psi1}).
The asymptotic results for small and large separation from the
wall, Eqs.(\ref{fasym1}) and (\ref{fasym2}), are denoted by broken lines.}
\caption{Plot of the rescaled one-loop correction to the 
counter ion density $h$, as defined by the Eqs.(\ref{hdef}) and 
(\ref{hsol}). One notes that the density change at the substrate
surface vanishes identically.}
\caption{ Plot of the one-loop prediction for
the rescaled density
$\bar{\rho}(z) = (\tilde{\ell}_B \mu^2/2) \rho(z)$
as a function of the rescaled distance from the wall
$z/\mu$, for four different values of the parameter
$\tilde{\ell}_B/4 \pi \mu$. The broken line
denotes the PB result;
the other three solid lines  are for $\tilde{\ell}_B/4 \pi \mu =1$,
$5$, and $10$, with the distance to the PB curve
increasing as $\tilde{\ell}_B/4 \pi \mu$
increases. }
\end{figure}


\begin{references}

\bibitem{Gouy}
G. Gouy, J. Phys. {\bf 9}, 457 (1910); Ann. Phys. {\bf 7}, 129 (1917).

\bibitem{Chapman}
D.L. Chapman, Philos. Mag. {\bf 25}, 475 (1913).

\bibitem{Stern}
O. Stern, Z. Elektrochem. {\bf 30}, 508 (1924).

\bibitem{Fuoss}
R.M. Fuoss, A. Katachalsky, and S. Lifson, Proc. Natl. Acad. Sci. U.S.A.
{\bf 37}, 579 (1951).

\bibitem{Andelman}
D. Andelman, Ch. 12 in {\em Handbook of
Biological Physics, Vol. 1}, R. Lipowsky and E. Sackmann, eds.,
(Elsevier,  1995).

\bibitem{Outh}
S. Levine and C.W. Outhwaite, J. Chem. Soc., Faraday Trans. 2 {\bf 74},
1670 (1978); C.W. Outhwaite and L.B. Bhuiyan, {\em ibid.}
{\bf 79}, 707 (1983).

\bibitem{Henderson}
D. Henderson, L.Blum, and W.R. Smith,
Chem. Phys. Lett. {\bf 63}, 381 (1979).

\bibitem{Forstmann}
P. Nielaba and F. Forstmann, Chem. Phys. Lett. {\bf 117}, 46 (1985);
B. D'Aguanno, P. Nielaba, T. Alts, and F. Forstmann {\bf 85}, 3476 (1986);
P. Nielaba, T. Alts, B. D'Aguanno, and F. Forstmann {\bf 34}, 1505 (1986).

\bibitem{Torrie}
G.M. Torrie and J.P. Valleau, Chem. Phys. Lett. {\bf 65}, 343 (1979);
J. Chem. Phys. {\bf 73}, 5807 (1980); J. Phys. Chem. {\bf 86}, 3251 (1982).

\bibitem{Megen}
W. van Megen and I Snook, J. Chem. Phys. {\bf 73}, 4656 (1980).

\bibitem{Svensson}
B. Svensson and B. J\"onsson, Chem. Phys. Lett. {\bf 108}, 580 (1984).

\bibitem{Svensson2}
B. Svensson, B. J\"onsson, and C.E. Woodward, J. Phys. Chem. {\bf 94},
2105 (1990).

\bibitem{Marcelo}
M. Lozada-Cassou, J. Chem. Phys. {\bf 80}, 3344 (1984).

\bibitem{Kjellander}
R. Kjellander and S. Mar\^celja, Chem. Phys. Lett. {\bf 112}, 49 (1984);
J. Chem. Phys. {\bf 82}, 2122 (1985).

\bibitem{Kjellander2}
R. Kjellander, T. Akesson, B. J\"onsson, and S. Mar\^celja,
J. Chem. Phys. {\bf 97}, 1424 (1992).

\bibitem{Jonsson}
B. J\"onsson, H. Wennerstr\"om, and B. Halle, J. Phys. Chem. {\bf 84},
2179 (1980).

\bibitem{Guldbrand}
L. Guldbrand, B. J\"onsson, H. Wennerstr\"om, and P. Linse,
J. Chem. Phys. {\bf 80}, 2221 (1984).

\bibitem{Attard}
P. Attard, Adv. Chem. Phys. {\bf XCII}, 1 (1996).

\bibitem{Podgornik}
R. Podgornik and B. Zeks, J. Chem. Soc. Faraday Trans. 2 {\bf 84}, 611 
(1988);
R. Podgornik, J. Phys. A {\bf 23}, 275 (1990).

\bibitem{Attard2}
P. Attard, D.J. Mitchell, and B.W. Ninham, J. Chem. Phys. {\bf 88}, 4987
(1988); {\em ibid.} {\bf 89}, 4358 (1988).

\bibitem{Schwinger}
J. Schwinger, Proc. Natl. Acad. Sci. {\bf 37}, 452 (1951);
{\bf 37}, 455 (1951); {\bf 44}, 956 (1958).

\bibitem{BLZ}
E. Br\'ezin, J.C. Le Guillou, and J. Zinn-Justin, in {\em Phase 
Transitions and Critical Phenomena}, Vol. VI, C. Domb and M.S. Green, eds. 
(Academic Press, N.Y., 1976).

\bibitem{Abramowitz}
{\em Handbook of Mathematical Functions},
M. Abramowitz and I.A. Stegun, eds. (Dover Publications, New York, 1972).

\bibitem{Jancovici}
B. Jancovici, J. Stat. Phys. {\bf 28}, 43 (1982).

\bibitem{Baus}
M. Baus, Mol. Phys. {\bf 48}, 347 (1983).

\bibitem{Carnie}
S.L. Carnie and D.Y. Chan, Mol. Phys. {\bf 51}, 1047 (1984).
\end{references}
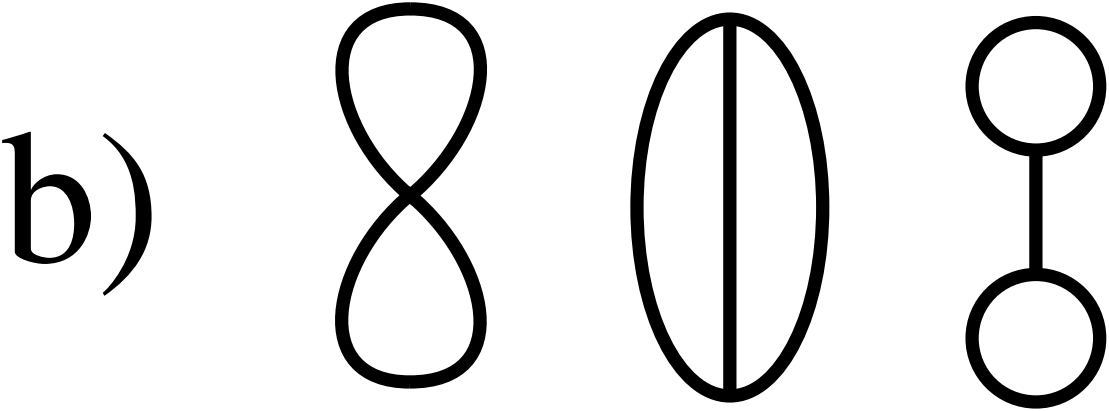
\end{document}